\documentclass[conference]{IEEEtran}
\ifCLASSINFOpdf
\else
   \usepackage{graphicx}
   \graphicspath{{../eps/}}
   \DeclareGraphicsExtensions{.eps}
\fi

\usepackage{amsmath}
\usepackage{amsfonts,amssymb}
\usepackage{amssymb}
\usepackage{wrapfig}
\usepackage{psfrag}
\usepackage{epstopdf}
\usepackage{cite}
\usepackage{graphicx}
\usepackage{subfigure}
\usepackage{threeparttable}
\usepackage{cases}
\usepackage{subeqnarray}
\usepackage{color}
\usepackage{underscore}
\usepackage{verbatim}
\usepackage{bm}
\usepackage{stfloats}
\newtheorem{theorem}{Theorem}
\newtheorem{lemma}{Lemma}

\usepackage{algorithm}
\usepackage{algorithmic}

\begin{document}

\title{How Does Performance Scale with Antenna Number for Extremely Large-Scale MIMO?}
%
%
%
\author{\IEEEauthorblockN{$\text{Haiquan~Lu}^*$ and $\text{Yong~Zeng}^*\dagger$}
\IEEEauthorblockA{*National Mobile Communications Research Laboratory, Southeast University, Nanjing 210096, China\\
$\dagger$Purple Mountain Laboratories, Nanjing 211111, China\\
Email: \{haiquanlu, yong_zeng\}@seu.edu.cn.}
}

\maketitle


\begin{abstract}
  Extremely large-scale multiple-input multiple-output (XL-MIMO) communications correspond to systems whose antenna size is so large that conventional assumptions, such as uniform plane wave (UPW) impingement, are no longer valid. This paper studies the channel modelling and performance analysis of XL-MIMO communication based on the generic spherical wavefront propagation model. First, for the single-user uplink/downlink communication with the optimal maximum ratio combining/transmission (MRC/MRT), we rigorously derive a new closed-form expression for the resulting signal-to-noise ratio (SNR), which includes the conventional SNR expression based on UPW assumption as a special case. Our result shows that instead of scaling linearly with the base station (BS) antenna number $M$, the SNR with the more generic spherical wavefront model increases with $M$ with diminishing return, governed by a new parameter called \emph{angular span}. One important finding from our derivation is the necessity to introduce a new distance criterion, termed \emph{critical distance}, to complement the classical Rayleigh distance for separating the near- and far-field propagation regions. While Rayleigh distance is based on the phase difference across array elements and hence depends on the \emph{electrical size} of the antenna, the critical distance cares about the amplitude/power difference and only depends on its \emph{physical size}. We then extend the study to the multi-user XL-MIMO communication system, for which we demonstrate that inter-user interference (IUI) can be mitigated not just by angle separation, but also by distance separation along the same direction. This offers one new degree of freedom (DoF) for interference suppression with XL-MIMO.
\end{abstract}

%

\IEEEpeerreviewmaketitle
\section{Introduction}
With the fifth-generation (5G) mobile communication networks being deployed worldwide, massive multiple-input multiple-output (MIMO) as a key enabling technology for significantly improving spectral efficiency has become a reality~\cite{marzetta2010noncooperative}. A typical 5G base station (BS) would be equipped with a massive MIMO array of 64 antenna elements~\cite{Zhang2020MultipleAT}. Looking ahead for beyond 5G (B5G) or sixth-generation (6G) era, there has been growing interest in pushing massive MIMO to another level by further increasing the antenna size/number drastically, for which concepts like ultra-massive MIMO (UM-MIMO), extremely large-scale MIMO (XL-MIMO), and extremely large aperture massive MIMO (xMaMIMO) have been proposed~\cite{akyildiz2016realizing,bjornson2019massive,decarvalho2020nonstationarities,amiri2018extremely}.

Compared to existing massive MIMO systems, several new channel characteristics emerge when moving towards XL-MIMO. In particular, the extremely large-scale antenna array at the BS, together with the continuous reduction of cell size, renders the users/scatterers less likely to be located in the far-field region, where conventional uniform plane wave (UPW) assumption is usually made for channel modelling and performance analysis. Note that the typical way for separating far-field versus radiative near-field regions is based on the classical Rayleigh distance ${r_{{\rm{Rayl}}}} = \frac{{2{D^2}}}{\lambda }$, where $D$ and $\lambda$ denote the physical dimension of the antenna array and signal wavelength, respectively~\cite{Payami2012Channel,zhou2015spherical,friedlander2019localization,balanis2016antenna}. More specifically, $r_{\rm{Rayl}}$ corresponds to the minimum link distance so that when the array is used for reception, the maximum phase difference of the received signals across the array elements is no greater than $\frac{\pi }{{\rm{8}}}$~\cite{balanis2016antenna,selvan2017fraunhofer}. In the far-field region with the link distance $r \geq r_{\rm{Rayl}}$, the signals can be well approximated as UPW, for which all array elements share identical signal amplitude and angle of arrival/departure (AoA/AoD) for each channel path. However, as the antenna size $D$ further increases so that the users/scatterers are located within the Rayleigh distance, near-field radiation with the more generic spherical wavefront needs to be considered to more accurately model both the signal phase and amplitude variations across array elements. Some preliminary efforts have been devoted towards this direction. In~\cite{zhou2015spherical}, the spherical wave channel under line-of-sight (LoS) conditions was proposed. In~\cite{friedlander2019localization}, the near-field array manifold based on the spherical wavefront model was derived to estimate the location of the near-field source. Apart from the planar and spherical wavefront models, an intermediate parabolic wave model was introduced in~\cite{le2019massive} to achieve a balance between model accuracy and complexity.

Besides spherical wavefront modelling, another new characteristic of XL-MIMO is the spatial channel non-stationarity in complex propagation environment. Specifically, different regions of the array may observe distinct propagation environment (e.g., different blockers and/or cluster sets), and hence exhibit different levels of received power. One approach for characterizing such spatial non-stationarity is via the concept of visibility region of the array~\cite{bjornson2019massive,decarvalho2020nonstationarities,amiri2018extremely,han2020channel}.

This paper focuses on the channel modelling and performance analysis of XL-MIMO communication by taking into account the generic spherical wavefront propagation, as opposed to the conventional UPW assumption. To gain useful insights, we first consider the single-user uplink/downlink communication with the optimal maximum ratio combining/transmission (MRC/MRT). A new closed-form expression for the resulting signal-to-noise ratio (SNR) is derived, which generalizes the conventional SNR expression that relies on the simplified UPW assumption. The derived result shows that instead of scaling linearly with the BS antenna number $M$, the resulting SNR based on the more general spherical wavefront model increases with $M$ with diminishing return, governed by a new parameter called \emph{angular span}, i.e., the angle formed by the two line segments connecting the user with both ends of the antenna array, as illustrated in Fig.~\ref{MultiusersystemModel}. Particularly, as $M \to \infty $, the SNR approaches to a constant value that depends on the user's projected distance to the antenna array. This is in a sharp contrast with the conventional UPW-based result that SNR grows linearly and unbounded with $M$, which is obviously impractical and misleading. One important finding from our derivation is the necessity to introduce a new distance partition criterion, termed \emph{critical distance}, to complement the classical Rayleigh distance for separating the near- and far-field regions. Different from the Rayleigh distance that is based on the phase difference across array elements and hence depends on the \emph{electrical size} of the antenna array, the critical distance cares about the amplitude/power difference, which only depends on the \emph{physical size} of the antenna array. We then extend our study to multi-user XL-MIMO communication, for which we demonstrate that inter-user interference (IUI) can be mitigated not just by angle separation as in conventional multi-antenna systems, but also by distance separation along the same direction. This thus provides a new degree of freedom (DoF) for interference suppression with XL-MIMO. Numerical results are presented to demonstrate the importance of proper spherical wavefront modelling for XL-MIMO communications, by showing the big performance deviation of the conventional UPW-based results from the true values.


\section{System Model}
 \begin{figure}[!t]
  \setlength{\abovecaptionskip}{-0.1cm}
  \setlength{\belowcaptionskip}{-0.3cm}
  \centering
  \centerline{\includegraphics[width=2.6in,height=2.67in]{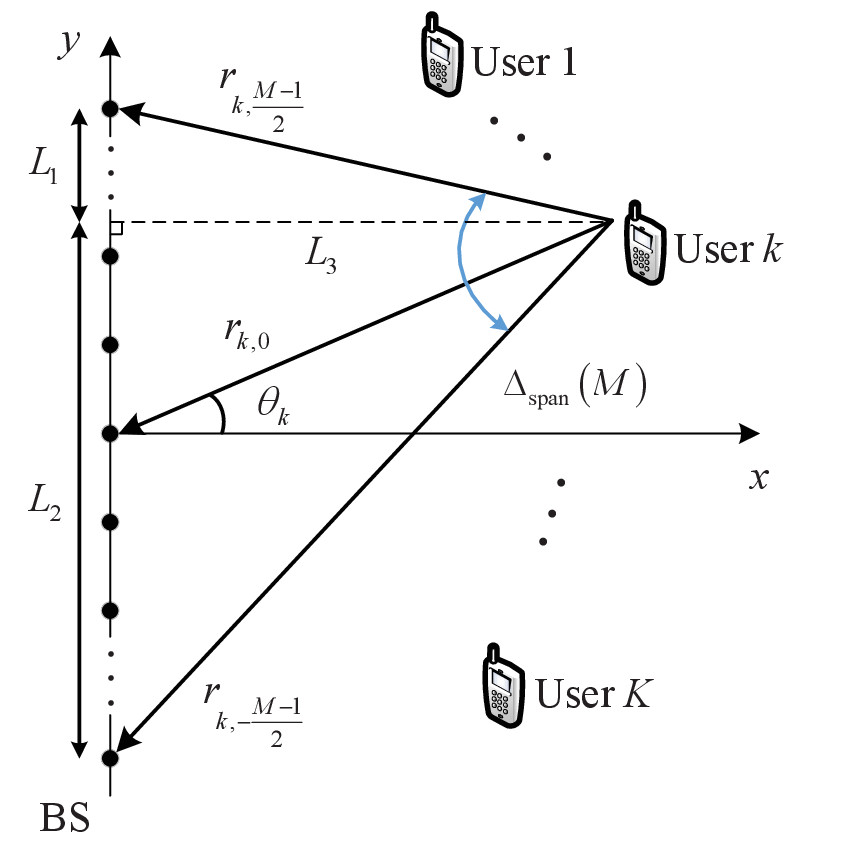}}
  \caption{Wireless communication with extremely large-scale MIMO.}
  \label{MultiusersystemModel}
  \vspace{-0.5cm}
  \end{figure}

As shown in Fig.~\ref{MultiusersystemModel}, we consider a wireless communication system, where a BS with an extremely large-scale antenna array serves $K$ users. For simplicity, we assume that the BS is equipped with an uniform linear array (ULA) with $M \gg 1$ elements, and each user has one single antenna. Without loss of generality, the ULA is placed along the $y$-axis and centered at the origin, and $M$ is an odd number for notational convenience. The location of the $m$th array element is ${{\bf{w}}_m} = {\left[ {0,md} \right]^T}$, where $m = 0, \pm 1, \cdots , \pm \left( {M - 1} \right)/2$, and $d \le \frac{\lambda }{2}$ is the antenna element separation with $\lambda$ denoting the signal wavelength. The location of user $k$ is denoted by ${{\bf{q}}_k} = {\left[ {{r_k}\cos {\theta _k},{r_k}\sin {\theta _k}} \right]^T},\;1 \le k \le K$, where ${r_k}$ denotes the distance between user $k$ and the center of the antenna array, and $\theta_k  \in \left[ { - \frac{\pi }{2},\frac{\pi }{2}} \right]$ is the angle of the line segment connecting to the user's location and the array center relative to the $x$-axis. The distance between user $k$ and the $m$th antenna is then given by
\begin{equation}\label{UserAndmthAntennaDistance}
{r_{k,m}} = \left\| {{{\bf{w}}_m} - {{\bf{q}}_k}} \right\| = {r_k}\sqrt {1 - 2m{\epsilon _k}\sin {\theta _k} + {m^2}\epsilon _k^2},
\end{equation}
where ${\epsilon _k} \buildrel \Delta \over = \frac{d}{{{r_k}}}$. Note that ${r_k} = {r_{k,0}}$ and since the array element separation $d$ is on the order of wavelength, in practice, we have $\epsilon_k\ll 1$.

By assuming that the communication links between the BS and users are dominated by the LoS component, the array response vector for user $k$ located at distance $r_k$ with angle $\theta_k$ can be expressed as
\begin{equation}\label{UserkArrayResponse}
  \setlength\abovedisplayskip{1pt}
 \setlength\belowdisplayskip{1pt}
{\small
\begin{aligned}
{\bf{a}}\left( {{r_k},{\theta _k}} \right) = \left[ {{a_{ - \frac{{M - 1}}{2}}}\left( {{r_k},{\theta _k}} \right), \cdots } \right.,{a_m}\left( {{r_k},{\theta _k}} \right) {\left. {, \cdots ,{a_{\frac{{M - 1}}{2}}}\left( {{r_k},{\theta _k}} \right)} \right]^T},
\end{aligned}}
\end{equation}
where ${a_m}\left( {{r_k},{\theta _k}} \right) = \frac{{\sqrt {{\beta _0}} }}{{{r_{k,m}}}}{e^{ - j\frac{{2\pi }}{\lambda }{r_{k,m}}}}$ with ${\beta _0}$ denoting the channel power at the reference distance ${d_0} = 1$~m. Note that the distance $r_{k,m}$ affects both the amplitude and phase of the array response in \eqref{UserkArrayResponse}.

For uplink communication{\footnote[1]{Our results can be easily extended to downlink communication.}}, the received signal at the BS can be expressed as
\begin{equation}\label{BSRaeceivedSignalforUserkWithoutMRC}
{{\bf{y}}_k} = {\bf{a}}\left( {{r_k},{\theta _k}} \right)\sqrt {{P_k}} {s_k} + \sum\limits_{i = 1,i \ne k}^K {{\bf{a}}\left( {{r_i},{\theta _i}} \right)\sqrt {{P_i}} {s_i}}  + {\bf{n}},
\end{equation}
where $P_k$ and $s_k, 1 \le k \le K$ are the transmit power and information-bearing signal of user $k$, respectively; ${\bf{n}} \sim {\cal C}{\cal N}\left( {0,{\sigma ^2}{{\bf{I}}_M}} \right)$ is the additive white Gaussian noise (AWGN). With linear receive beamforming ${{\bf{v}}_k} \in {{\mathbb{C}}^{M \times 1}}$ applied for user $k$, the resulting signal for user $k$ can be expressed as
\begin{equation}\label{BSRaeceivedSignalforUserk}
\begin{aligned}
{y_k} =& {\bf{v}}_k^H{\bf{a}}\left( {{r_k},{\theta _k}} \right)\sqrt {{P_k}} {s_k} +  \\
&{\bf{v}}_k^H\sum\limits_{i = 1,i \ne k}^K {{\bf{a}}\left( {{r_i},{\theta _i}} \right)\sqrt {{P_i}} {s_i}}  + {\bf{v}}_k^H{\bf{n}}.
\end{aligned}
\end{equation}
The received signal-to-interference-plus-noise ratio (SINR) for user $k$ is then given by
\begin{equation}\label{ReceivedSINRForkthUser}
{\gamma _k} = \frac{{{P_k}{{\left| {{\bf{v}}_k^H{\bf{a}}\left( {{r_k},{\theta _k}} \right)} \right|}^2}}}{{\sum\limits_{i = 1,i \ne k}^K {{P_i}{{\left| {{\bf{v}}_k^H{\bf{a}}\left( {{r_i},{\theta _i}} \right)} \right|}^2}}  + {{\left\| {{{\bf{v}}_k}} \right\|}^2}{\sigma ^2}}},\;1 \le k \le K.
\end{equation}
The achievable sum rate in bits/second/Hz (bps/Hz) is
\begin{equation}\label{SumRate}
{R_{{\rm{sum}}}} = \sum\limits_{k = 1}^K {{{\log }_2}\left( {1 + {\gamma _k}} \right)}.
\end{equation}
\section{Single-User Extremely Large-Scale Antenna Communication}\label{Single User System}
To gain some insights, we first focus on the special case of single-user system, i.e., $K=1$, for which user index $k$ is suppressed in this section. Due to the absence of IUI, it is known that the low-complexity MRC beamformer ${\bf{v}} = \frac{{{\bf{a}}\left( {r,\theta } \right)}}{{\left\| {{\bf{a}}\left( {r,\theta } \right)} \right\|}}$ is optimal. It follows from \eqref{UserAndmthAntennaDistance}, \eqref{UserkArrayResponse} and \eqref{ReceivedSINRForkthUser} that the resulting SNR can be written as
\begin{equation}\label{SNRExpressionAtBS}
{\gamma_{\rm SW}}  = \frac{{P{{\left\| {{\bf{a}}\left( {{r},\theta } \right)} \right\|}^2}}}{{{\sigma ^2}}} = \bar P\sum\limits_{m =  - \frac{{M - 1}}{2}}^{\frac{{M - 1}}{2}} {\frac{{{\beta _{\rm{0}}}{\rm{/}}r^2}}{{1 - 2m\epsilon \sin \theta  + {m^2}{\epsilon ^2}}}},
\end{equation}
where $\bar P \buildrel \Delta \over = \frac{P}{{{\sigma ^2}}}$ is the transmit SNR, and $\epsilon  = \frac{d}{r} \ll 1$. Note that the subscript ${\left(  \cdot  \right)_{{\rm{SW}}}}$ signifies that \eqref{SNRExpressionAtBS} is based on the generic spherical wave channel model as in \eqref{UserkArrayResponse}.

\begin{theorem} \label{SNRapproximationTheorem}
For single-user large-scale antenna communication with the optimal MRC/MRT beamforming, the resulting SNR can be expressed in closed-form as
\begin{equation}\label{ApproxiamtionSNR}
{\gamma_{\rm SW}} {\rm{ = }}\bar P\frac{{{\beta _{\rm{0}}}}}{{dr\cos \theta }}{\Delta _{{\rm{span}}}}\left( M \right),
\end{equation}
where ${\Delta _{{\rm{span}}}}\left( M \right) \buildrel \Delta \over = \arctan \left( {\frac{{Md}}{{{\rm{2}}{r}\cos \theta }} - \tan \theta } \right) + \arctan \left( {  \frac{{Md}}{{{\rm{2}}{r}\cos \theta }} + \tan \theta } \right)$.
\end{theorem}

 \begin{IEEEproof}
 Please refer to Appendix A.
 \end{IEEEproof}

 Theorem~\ref{SNRapproximationTheorem} shows that with the more generic spherical wavefront channel model in \eqref{UserkArrayResponse} (as opposed to conventional UPW assumption), the resulting SNR scales with the antenna number $M$ according to a new parameter ${\Delta _{{\rm{span}}}}\left( M \right)$, which we term as the \emph{angular span}.
 As illustrated in Fig.~\ref{MultiusersystemModel}, the angular span of a user is simply the angle formed by the two line segments connecting the user with both ends of the antenna array, i.e., ${\Delta _{{\rm{span}}}}\left( M \right) = \arctan \frac{{{L_1}}}{{{L_3}}} + \arctan \frac{{{L_2}}}{{{L_3}}} = \arctan \left( {\frac{{\frac{{ Md}}{{\rm{2}}} - {r}\sin \theta }}{{{r}\cos \theta }}} \right) + \arctan \left( {\frac{{\frac{{M d}}{{\rm{2}}} + {r}\sin \theta }}{{{r}\cos \theta }}} \right)$. It can be shown that as $M \to \infty$, ${\Delta _{{\rm{span}}}}\left( M \right) \to \pi $. This thus leads to
 \begin{equation}\label{limitValueofSNR}
\mathop {\rm lim}\limits_{M \to \infty } {\gamma _{{\rm{SW}}}} = \bar P\frac{{{\beta _0}\pi }}{{d{r}\cos \theta }},
\end{equation}
 which is a constant that depends on the user's projected distance to the ULA, i.e., ${r\cos \theta }$. By contrast, with the conventional UPW assumption, it is known that the resulting SNR with the optimal MRC/MRT beamforming is
 \begin{equation}\label{SNRConventionalUPW}
{\gamma _{{\rm{UPW}}}} = \bar P\frac{{M{\beta _{\rm{0}}}}}{{r^2}},
 \end{equation}
 which increases linearly with $M$ and grows unbounded when $M \to \infty$. This result is obviously impractical and misleading, and the fundamental reason lies in that the UPW assumption is no longer valid as $M$ goes large. This issue is resolved in our newly derived expression \eqref{ApproxiamtionSNR} based on the more generic spherical wavefront channel model \eqref{UserkArrayResponse}. Another important observation is that while ${\gamma _{{\rm{UPW}}}}$ only depends on the distance $r$, ${\gamma_{\rm SW}}$ depends on both distance $r$ and direction $\theta$.

  Furthermore, a closer look at the angular span in Fig.~\ref{MultiusersystemModel} leads to an alternative expression of ${\Delta _{{\rm{span}}}}\left( M \right)$, as shown below.
  \begin{lemma} \label{angularSpanExpression}
  The angular span of a user may also be expressed as
  \begin{equation}
  \setlength\abovedisplayskip{1pt}
  \setlength\belowdisplayskip{1pt}
  {\small
  \begin{aligned}
  {\Delta _{{\rm{span}}}}\left( M \right) = \arctan \left( {\frac{{\frac{{Md}}{2}\cos \theta }}{{{r} - \frac{{Md}}{2}\sin \theta }}} \right) + \arctan \left( {\frac{{\frac{{Md}}{2}\cos \theta }}{{{r} + \frac{{Md}}{2}\sin \theta }}} \right).
  \end{aligned}}
  \end{equation}
  \end{lemma}
\begin{IEEEproof}
This result can be obtained by an alternative partition of the angular span into two angles with an auxiliary lie segment connecting the origin and user location. The details are omitted for brevity.
\end{IEEEproof}

Lemma \ref{angularSpanExpression} is useful for showing the following result:
\begin{lemma} \label{GeneralExpressionlemma}
When $r \gg \frac{{Md}}{2}$, $\gamma_{\rm SW}$ in \eqref{ApproxiamtionSNR} reduces to
\begin{equation}
\gamma_{\rm SW}  \approx {\gamma _{{\rm{UPW}}}} = \bar P\frac{{M{\beta _0}}}{{r^2}}.
\end{equation}
\end{lemma}
\begin{IEEEproof}
 Please refer to Appendix B.
\end{IEEEproof}

Lemma \ref{GeneralExpressionlemma} shows that the new closed-form SNR expression in Theorem \ref{SNRapproximationTheorem} generalizes the conventional expression \eqref{SNRConventionalUPW} that relies on the UPW assumption, since the latter is included as a special case of \eqref{ApproxiamtionSNR}.

 \begin{figure}[!t]
  \setlength{\abovecaptionskip}{-0.05cm}
  \setlength{\belowcaptionskip}{-0.2cm}
  \centering
  \centerline{\includegraphics[width=3.5in,height=2.625in]{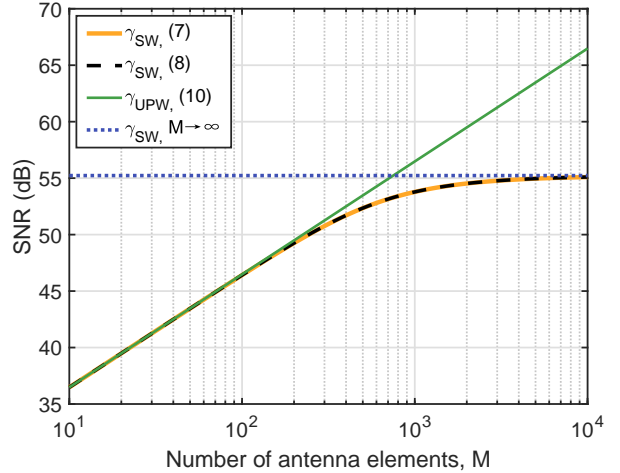}}
  \caption{SNR versus antenna number $M$ with UPW versus spherical wavefront models.}
  \label{exactApproximateLimitSNR}
  \vspace{-0.6cm}
  \end{figure}

 Fig.~\ref{exactApproximateLimitSNR} plots the SNR $\gamma_{\rm SW}$ and $\gamma_{\rm UPW}$ versus the number of antenna elements $M$ for one example scenario. For $\gamma_{\rm SW}$, the results based on both the summation in \eqref{SNRExpressionAtBS} and the closed-form expression \eqref{ApproxiamtionSNR} are shown, together with the SNR limit in \eqref{limitValueofSNR}. The location of the user is ${{\bf{q}}} = {\left[ {15,0} \right]^T}$~m. The carrier frequency is 2.4~GHz, and the antenna separation is $d = \frac{\lambda }{2} = {\rm{0.0628}}$~m. The received SNR at the reference distance is $\bar P{\beta _{\rm{0}}} = 50$~dB. It is firstly observed that our derived closed-form SNR expression \eqref{ApproxiamtionSNR} perfectly matches with the summation in \eqref{SNRExpressionAtBS}, which validates Theorem~\ref{SNRapproximationTheorem}. It is also observed that for relatively small $M$, $\gamma_{\rm SW}$ in \eqref{ApproxiamtionSNR} matches with $\gamma_{\rm UPW}$ in \eqref{SNRConventionalUPW}, which is in accordance with Lemma~\ref{GeneralExpressionlemma}. However, as $M$ increases further as in XL-MIMO regime, the two expressions exhibit drastically different scaling laws with $M$, i.e., approaching a constant value versus increasing linearly and unbounded. This demonstrates the importance of properly modelling the spherical wave nature in XL-MIMO systems, as in Theorem~\ref{SNRapproximationTheorem}.

 \begin{figure}[!t]
  \setlength{\abovecaptionskip}{-0.05cm}
  \setlength{\belowcaptionskip}{-0.2cm}
  \centering
  \centerline{\includegraphics[width=3.5in,height=2.625in]{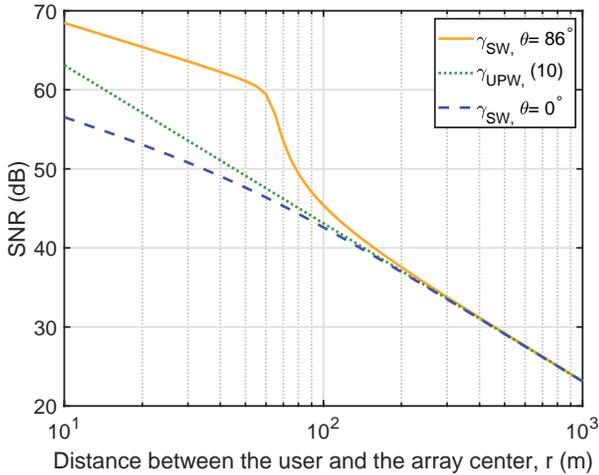}}
  \caption{SNR versus link distance $r$ with UPW versus spherical wavefront models.}
  \label{fixMDecreaseRateComparison}
  \vspace{-0.6cm}
  \end{figure}

 Fig.~\ref{fixMDecreaseRateComparison} shows the SNR $\gamma_{\rm SW}$ in \eqref{ApproxiamtionSNR} and $\gamma_{\rm UPW}$ in \eqref{SNRConventionalUPW} versus the link distance $r$ for two different user directions $\theta {\rm{ = }}{{\rm{0}}^ \circ }$ and ${\rm{8}}{{\rm{6}}^ \circ }$, with the antenna number fixed to $M=2048$. While $\gamma_{\rm UPW}$ is independent of $\theta$, the difference of $\gamma_{\rm SW}$ for the two considered $\theta$ values can be up to 10~dB for moderate link distance $r$. It is also observed that depending on $\theta$, the conventional UPW expression $\gamma_{\rm UPW}$ may either over-estimate (as for $\theta=0^ \circ$) or under-estimate (as for $\theta=86^ \circ$) the true value. As $r$ increases beyond certain limit, $\gamma_{\rm SW}$ merges with $\gamma_{\rm UPW}$, as predicted in Lemma~\ref{GeneralExpressionlemma}.

\begin{figure}[!t]
  \setlength{\abovecaptionskip}{-0.05cm}
  \setlength{\belowcaptionskip}{-0.4cm}
  \centering
  \centerline{\includegraphics[width=3.5in,height=2.625in]{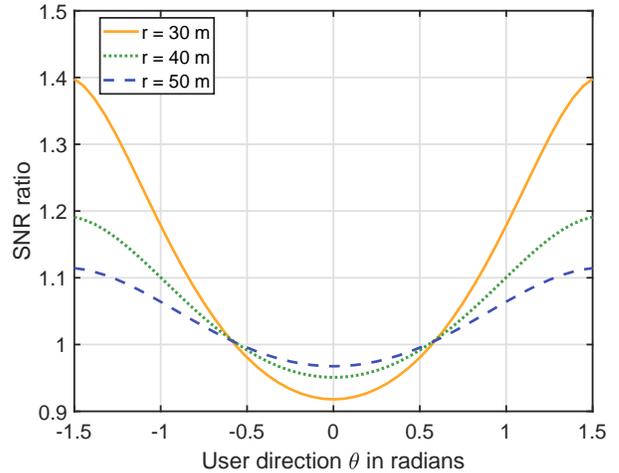}}
  \caption{SNR ratio versus user direction.}
  \label{RatioVersusTheta}
  \vspace{-0.5cm}
  \end{figure}

To further characterize the difference between $\gamma_{\rm SW}$ and $\gamma_{\rm UPW}$, we define the following SNR ratio
\begin{equation}\label{RatioBetweenGammaAndGammaUPW}
\setlength\abovedisplayskip{1pt}
\setlength\belowdisplayskip{1pt}
\Gamma \left( {r,\theta ,M} \right) \buildrel \Delta \over = \frac{{{\gamma _{{\rm{SW}}}}}}{{{\gamma _{{\rm{UPW}}}}}} = \frac{{r{\Delta _{{\rm{span}}}}\left( M \right)}}{{Md\cos \theta }}.
\end{equation}
For the special case when the user is located at the $x$-axis or $y$-axis, i.e., $\theta  = 0$ or $\theta  =  \pm \frac{\pi }{2}$, ${\gamma _{{\rm{SW}}}}$ in \eqref{ApproxiamtionSNR} reduces to
  \begin{equation}\label{theta=0/+-pi/2ApproxiamtionSNR}
    \setlength\abovedisplayskip{1pt}
  \setlength\belowdisplayskip{1pt}
  {\gamma _{{\rm{SW}}}} {\rm{ = }}\left\{ \begin{array}{l}
  \bar P\frac{{2{\beta _{\rm{0}}}}}{{d{r}}}\arctan \left( {\frac{{Md}}{{2{r}}}} \right),\ {\rm{ if }}\ \theta  = 0,\\
  \vspace{-0.2cm}
  \\
  \bar P\frac{{M{\beta _{\rm{0}}}}}{{r^2 - \frac{{{M^2}{d^2}}}{4}}},\ \ \ \ \ \ \ \ \ \ \ {\rm{  if }}\ \theta  =  \pm \frac{\pi }{2}\  {\rm and}\  r > \frac{Md}{2}.
  \end{array} \right.
  \end{equation}
As a result, for $\theta = 0$, we have $\Gamma \left( {r,\theta  = 0,M} \right) = \frac{{\arctan \left( {\frac{{Md}}{{2r}}} \right)}}{{\frac{{Md}}{{2r}}}} < 1$, since $\arctan x < x$ for $x > 0$. In other words, for users located at the boresight of the antenna array with $\theta=0$, the conventional SNR expression based on UPW assumption always over-estimates the true result. On the other hand, for $\theta  =  \pm \frac{\pi }{2}$ and $r > \frac{Md}{2}$, we have $\Gamma \left( {r,\theta  =  \pm \frac{\pi }{2},M} \right) = \frac{{{r^2}}}{{{r^2} - \frac{{{M^2}{d^2}}}{4}}} > 1$, which means that the conventional SNR expression always under-estimates the true result for users located at the $y$-axis with $r > \frac{Md}{2}$. While it is challenging to draw similar conclusion for other $\theta$ angles theoretically, the SNR ratio can be drawn numerically versus $\theta$ for different $r$, as shown in Fig.~\ref{RatioVersusTheta}, where the number of antenna elements is $M=512$. It is observed that the SNR ratio first decreases and then increases within the interval $\theta  \in \left[ { - \frac{\pi }{2},\frac{\pi }{2}} \right]$, and the smallest ratio occurs at $\theta = 0$ for all three setups. The result shows that the conventional UPW-based SNR expression deviates significantly from the true value, especially for users located away from the antenna boresight (i.e., relatively large $\left| \theta  \right|$).

A closer look at Lemma~\ref{GeneralExpressionlemma}, together with Fig.~\ref{exactApproximateLimitSNR} and \ref{fixMDecreaseRateComparison}, reveals that the criterion whether $\gamma_{\rm SW}$ merges with $\gamma_{\rm UPW}$ only depends on the comparison of the link distance $r$ and the array's \emph{physical} dimension $D = \left( {M - 1} \right)d \approx Md$, regardless of the signal wavelength $\lambda$. This is quite different from the conventional way of separating the near- and far-field regions, which is based on the Rayleigh distance ${r_{{\rm{Rayl}}}} = \frac{{2{D^2}}}{\lambda }$ that depends on the electrical dimension~\cite{Payami2012Channel,zhou2015spherical,friedlander2019localization,balanis2016antenna}. The fundamental reason behind such a deviation lies in that Rayleigh distance merely concerns the maximum tolerable \emph{phase difference} across array elements, while irrespective of the amplitude/power difference. However, as shown in \eqref{UserkArrayResponse}, the distance $r$ impacts the channel via both the phase and amplitude. In particular, for the single-user multi-antenna communication with the optimal MRC/MRT beamforming where the signal phases are appropriately aligned, it is the amplitude variations across different elements that affect the eventual SNR, as evident from \eqref{SNRExpressionAtBS}. As a consequence, the conventional way of distinguishing near- and far-field regions based on $r_{\rm Rayl}$ is insufficient. Instead, a more refined link distance partition that takes into account both the amplitude and phase differences across array elements is needed. To this end, we introduce a new distance $r_{\rm Critical}$, termed \emph{critical distance}, for which the power ratio between the weakest and strongest array elements is no smaller than a certain threshold $\alpha$, say $\alpha = 80\% $. It can be shown that for a user with link distance $r$ and angle $\theta$, the power ratio, denoted as $\eta \left( {r,\theta } \right)$, can be expressed as
\begin{equation}\label{powerRatioExpression}
\setlength\abovedisplayskip{1pt}
\setlength\belowdisplayskip{1pt}
\eta \left( {r,\theta } \right) = \left\{ \begin{array}{l}
{\frac{{{r^2}{{\cos }^2}\theta }}{{{r^2}{{\cos }^2}\theta  + {{\left( {r\left| {\sin \theta } \right| + \frac{D}{2}} \right)}^2}}},\ \ {\rm{if}}\ r\left| {\sin \theta } \right| \le \frac{D}{2}},\\
\vspace{-0.2cm}
\\
{\frac{{{r^2}{{\cos }^2}\theta  + {{\left( {r\left| {\sin \theta } \right| - \frac{D}{2}} \right)}^2}}}{{{r^2}{{\cos }^2}\theta  + {{\left( {r\left| {\sin \theta } \right| + \frac{D}{2}} \right)}^2}}}},\ \ {\rm{otherwise }}.
\end{array} \right.
\end{equation}
$r_{\rm Critical}$ is thus defined as the minimum link distance $r$ such that $\eta \left( {r,\theta } \right) \ge \alpha ,\forall \theta  \in \left[ { - \frac{\pi }{2},\frac{\pi }{2}} \right]$. It can be shown that for any given $r$, $\eta \left( {r,\theta } \right)$ has the minimal value when
$\theta  = \pm \frac{\pi }{2}$. By substituting $\theta  =\pm \frac{\pi }{2}$ into \eqref{powerRatioExpression}, to ensure $\frac{{{{\left( {r - \frac{D}{2}} \right)}^2}}}{{{{\left( {r + \frac{D}{2}} \right)}^2}}} \ge 80\% $, we should have $r \ge \frac{{9 + \sqrt {80} }}{2}D$, i.e., ${r_{{\rm{Critical}}}} \approx 9D$, which only depends on the physical dimension of the antenna array. For $M\gg 1$, it readily follows that $r_{\rm Critical} < r_{\rm Rayl}$. As a result, a more refined partition of the link distance follows, as illustrated in Fig.~\ref{distancePartition}:
 \begin{figure}[!t]
  \setlength{\abovecaptionskip}{-0.2cm}
  \setlength{\belowcaptionskip}{-0.4cm}
  \centering
  \centerline{\includegraphics[width=3.3in,height=0.7in]{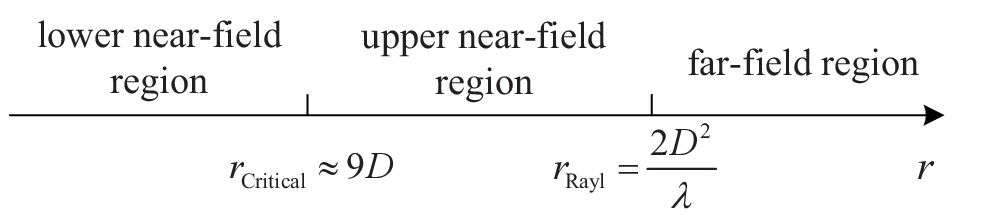}}
  \caption{An illustration of the refined link distance partition.}
  \label{distancePartition}
  \vspace{-0.5cm}
  \end{figure}

1) ${r_k} \ge {r_{{\rm{Rayl}}}}$: In this case, both the amplitude and phase differences across array elements are small, and UPW assumption gives good approximation. The model in \eqref{UserkArrayResponse} simplifies to ${a_m}\left( {{r_k},{\theta _k}} \right) = \frac{{\sqrt {{\beta _0}} }}{{{r_k}}}{e^{ - j\frac{{2\pi }}{\lambda }\left( {{r_k} - md\sin {\theta _k}} \right)}}$.

2) ${r_{{\rm{Critical}}}} \le {r_k} < {r_{{\rm{Rayl}}}}$: In this case, the amplitude difference is relatively small but the phase difference is large across elements. Therefore, the channel model in \eqref{UserkArrayResponse} can be simplified to ${a_m}\left( {{r_k},{\theta _k}} \right) = \frac{{\sqrt {{\beta _0}} }}{{{r_k}}}{e^{ - j\frac{{2\pi }}{\lambda }{r_{k,m}}}}$, and we term this region as \emph{upper near field region}.

3) ${r_k} < {r_{{\rm{Critical}}}}$: In this case, both amplitude and phase variations are large, and no simplification to \eqref{UserkArrayResponse} can be made, i.e., ${a_m}\left( {{r_k},{\theta _k}} \right) = \frac{{\sqrt {{\beta _0}} }}{{{r_{k,m}}}}{e^{ - j\frac{{2\pi }}{\lambda }{r_{k,m}}}}$. We term this region as \emph{lower near-field region}.

\section{Multi-user XL-MIMO Communication}\label{Multiuser System}
Next, we consider the more general multi-user XL-MIMO communications. For simplicity, we focus on the low-complexity MRC receive beamforming with ${{\bf{v}}_k} = \frac{{{\bf{a}}\left( {{r_k},{\theta _k}} \right)}}{{\left\| {{\bf{a}}\left( {{r_k},{\theta _k}} \right)} \right\|}}$, $\forall  k$, for which the SINR in \eqref{ReceivedSINRForkthUser} reduces to
\begin{equation}\label{MRCReceivedSINRForkthUser}
 \setlength\abovedisplayskip{1pt}
 \setlength\belowdisplayskip{1pt}
\begin{aligned}
{\gamma _k} &= \frac{{{P_k}{{\left\| {{\bf{a}}\left( {{r_k},{\theta _k}} \right)} \right\|}^2}}}{{\sum\limits_{i = 1,i \ne k}^K {{P_i}\frac{{{{\left| {{{\bf{a}}^H}\left( {{r_k},{\theta _k}} \right){\bf{a}}\left( {{r_i},{\theta _i}} \right)} \right|}^2}}}{{{{\left\| {{\bf{a}}\left( {{r_k},{\theta _k}} \right)} \right\|}^2}}}}  + {\sigma ^2}}}\\
 & = \frac{{{P_k}{{\left\| {{\bf{a}}\left( {{r_k},{\theta _k}} \right)} \right\|}^2}}}{{\sum\limits_{i = 1,i \ne k}^K {{P_i}{\rho _{ki}}} {{\left\| {{\bf{a}}\left( {{r_i},{\theta _i}} \right)} \right\|}^2} + {\sigma ^2}}},\;1 \le k \le K.
\end{aligned}
\end{equation}
where ${\rho _{ki}} \buildrel \Delta \over = \frac{{{{\left| {{{\bf{a}}^H}\left( {{r_k},{\theta _k}} \right){\bf{a}}\left( {{r_i},{\theta _i}} \right)} \right|}^2}}}{{{{\left\| {{\bf{a}}\left( {{r_k},{\theta _k}} \right)} \right\|}^2}{{\left\| {{\bf{a}}\left( {{r_i},{\theta _i}} \right)} \right\|}^2}}}$ denotes the correlation coefficient between user $k$ and $i$. It is noted that different from the conventional modelling based on UPW assumption where the correlation coefficient only depends on the user directions, that for XL-MIMO depends on both the link distances and angle separations.

\begin{figure}[!t]
  \setlength{\abovecaptionskip}{-0.1cm}
  \setlength{\belowcaptionskip}{-0.4cm}
  \centering
  \centerline{\includegraphics[width=3.5in,height=2.625in]{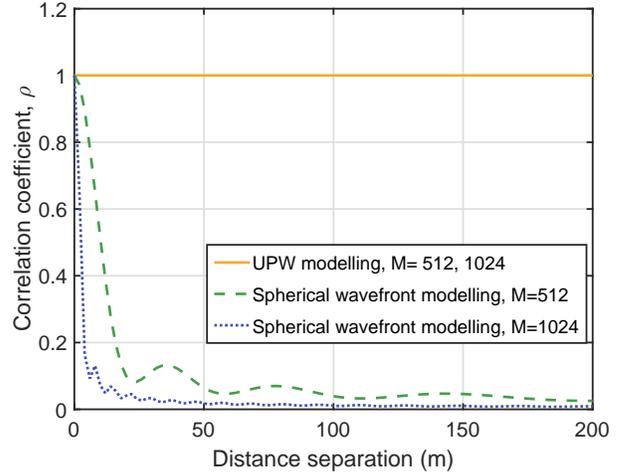}}
  \caption{Correlation coefficient versus the distance separation for UPW versus spherical wavefront models.}
  \label{correlationCoefficientVersusDistance}
  \vspace{-0.4cm}
  \end{figure}

Fig.~\ref{correlationCoefficientVersusDistance} shows the correlation coefficient ${\rho _{12}}$ versus the distance separation $\left| {{r_1} - {r_2}} \right|$ for the conventional UPW and the generic modellings, with the antenna number $M=512$, $1024$, respectively. We assume that user 1 is fixed at ${{\bf{q}}_{\rm{1}}} = \left[ {150,0} \right]$~m, and the two user directions are ${\theta _1} = {\theta _2} = 0$. It is observed that as the distance separation increases, the correlation coefficient based on the generic modelling decreases in general, as expected, while that based on the conventional modelling is always equal to one. In addition, it is also observed that the increase of antenna number $M$ further helps to reduce the correlation. This implies that the IUI could be mitigated not just by the direction separation as in the conventional UPW modelling, but also by the distance separation along the same direction. Such result shows that there exists a new DoF for interference suppression in XL-MIMO communications.

 \begin{figure}[!t]
  \setlength{\abovecaptionskip}{-0.05cm}
  \setlength{\belowcaptionskip}{-0.4cm}
  \centering
  \centerline{\includegraphics[width=3.5in,height=2.625in]{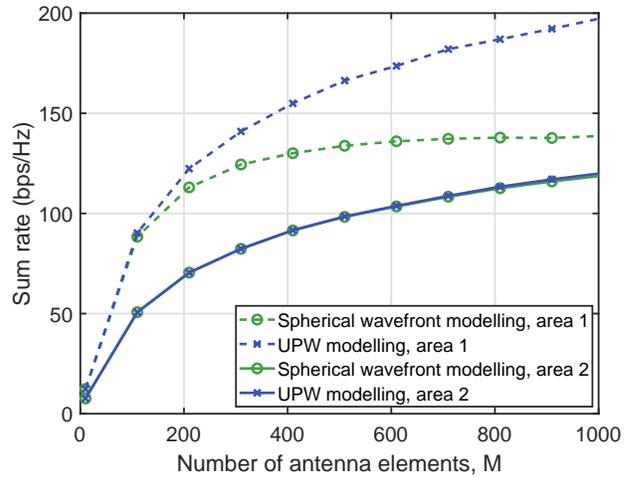}}
  \caption{Sum rate versus the antenna number for UPW versus spherical wavefront models.}
  \label{sumRateVersusNumberofAntennas}
  \vspace{-0.6cm}
  \end{figure}
Fig.~\ref{sumRateVersusNumberofAntennas} shows the sum rate of the multi-user XL-MIMO communication versus the antenna number $M$ with $K=10$ users. We assume that the users are randomly distributed in two areas with area 1 given by ${r_k} \in \left[ {100,200} \right]$~m, ${\theta _k} \in \left[ {-\frac{\pi }{4},\frac{\pi }{4}} \right]$, and area 2 ${r_k} \in \left[ {1000,1200} \right]$~m, ${\theta _k} \in \left[ {-\frac{\pi }{4},\frac{\pi }{4}} \right]$, $\forall k$. The reference SNR of all users is ${{\bar P}_k}{\beta _{\rm{0}}}{\rm{ = }}\frac{{{P_k}{\beta _{\rm{0}}}}}{{{\sigma ^2}}} = 50$~dB, $\forall k$. It is observed that for small $M$, the conventional modelling based on UPW assumption matches with the generic spherical wavefront modelling for both areas considered. Furthermore, for area 2 where users are located far away from the BS, both UPW and the proposed generic modelling give the same results across all $M$ values considered. By contrast, for area 1, with moderately large $M$, the conventional UPW modelling significantly over-estimates the true result. This is due to the fact that with the conventional UPW modelling, each user is treated for having only one single common AoA across all array elements, which thus exaggerates the angle separation for different users when each of them in fact has different AoAs with respect to different array regions for XL-MIMO. As a consequence, the IUI is severely under-estimated with the conventional UPW modelling, and hence the sum rate is over-estimated. This again shows the importance of appropriate spherical wavefront modelling for XL-MIMO communications.

\section{Conclusion}
This paper studied the channel modelling and performance analysis of XL-MIMO communication based on the generic spherical wavefront model. For single-user communication with the optimal MRC/MRT beamforming, our newly derived closed-from expression shows that the SNR with the more generic spherical wavefront modelling increases with the antenna number $M$ with diminishing return, governed by a new parameter called angular span. This is in a sharp contrast to the conventional UPW modelling where SNR scales linearly with $M$. Furthermore, we proposed a new distance criterion, termed critical distance, to complement the classical Rayleigh distance for separating the near- and far-field propagation regions. Then for the multi-user XL-MIMO communication system, it was demonstrated that the IUI could be mitigated not just by angle separation as in the conventional modelling, but also by distance separation along the same direction. Numerical results showed the importance of proper spherical wavefront modelling for XL-MIMO communications.
\begin{appendices}

\section{Proof of Theorem 1}
 \begin{figure}[!t]
  \setlength{\abovecaptionskip}{-0.2cm}
  \setlength{\belowcaptionskip}{-0.4cm}
  \centering
  \centerline{\includegraphics[width=3.5in,height=0.95in]{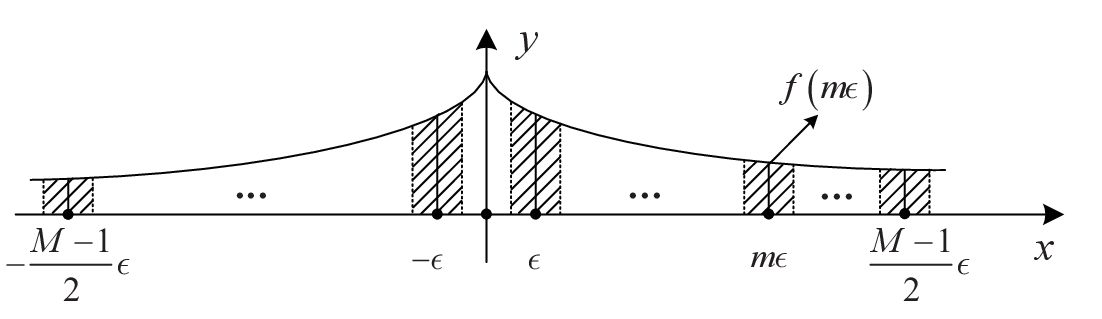}}
  \caption{An illustration of approximating the summation \eqref{SNRExpressionAtBS} by an integral.}
  \label{integralillustration}
  \vspace{-0.2cm}
  \end{figure}
We first consider the case when $\theta  \ne  \pm \frac{\pi }{2}$ and define the function $f\left( x \right) \buildrel \Delta \over = \frac{{{\beta _{\rm{0}}}{\rm{/}}r^2}}{{1 - 2x\sin \theta  + {x^2}}}$, which is a continuous function over the interval $x \in \left[ { - \frac{M}{2}\epsilon ,\frac{M}{2}\epsilon } \right]$. The interval is partitioned into $M$ subintervals each of equal length $\epsilon$, as shown in Fig.~\ref{integralillustration}. Since $\epsilon \ll 1$ in practice, we have $f(x) \approx f\left( {m\epsilon } \right)$, $ \forall x \in \left[ {\left( {m - \frac{1}{2}} \right)\epsilon ,\left( {m + \frac{1}{2}} \right)\epsilon } \right]$. Based on the concept of definite integral, we have
\begin{equation}\label{approximateintegral}
 \setlength\abovedisplayskip{1pt}
 \setlength\belowdisplayskip{1pt}
\sum\limits_{m =  - \frac{{M - 1}}{2}}^{\frac{{M - 1}}{2}} {f\left( {m\epsilon } \right)\epsilon }  \approx \int_{ - \frac{M}{2}\epsilon }^{\frac{M}{2}\epsilon } {f\left( x \right)} dx,
\end{equation}
where \eqref{approximateintegral} holds since $\epsilon  \ll 1$. By substituting ${f\left( x \right)}$ into \eqref{approximateintegral}, we have
\begin{equation}\label{approximateintegralDerivation}
 \setlength\abovedisplayskip{1pt}
 \setlength\belowdisplayskip{1pt}
\hspace{-0.55cm}
{\small
\begin{aligned}
&\sum\limits_{m =  - \frac{{M - 1}}{2}}^{\frac{{M - 1}}{2}} {\frac{{{\beta _{\rm{0}}}{\rm{/}}r^2}}{{1 - 2m\epsilon \sin \theta  + {m^{\rm{2}}}{\epsilon ^{\rm{2}}}}}}  \approx \frac{{\rm{1}}}{\epsilon }\int_{ - \frac{M}{2}\epsilon }^{\frac{M}{2}\epsilon } {\frac{{{\beta _{\rm{0}}}{\rm{/}}r^2}}{{1 - 2x\sin \theta  + {x^2}}}} dx\\
&\mathop = \limits^{\left( a \right)} \left. {\frac{{{\beta _{\rm{0}}}}}{{\epsilon r^2\cos \theta }}\arctan \left( {\frac{{x - \sin \theta }}{{\cos \theta }}} \right)} \right|_{ - \frac{M}{2}\epsilon }^{\frac{M}{2}\epsilon }\\
& = \frac{{{\beta _{\rm{0}}}}}{{dr\cos \theta }}\left[ {\arctan \left( {\frac{{Md}}{{2r\cos \theta }} - \tan \theta } \right) + } \right.\left. {\arctan \left( {\frac{{Md}}{{2r\cos \theta }} + \tan \theta } \right)} \right],
\end{aligned}}
\end{equation}
where ${\left( a \right)}$ follows from the integral formula 2.103 in~\cite{gradshteyn2014table}, i.e., $\int {\frac{{dx}}{{A + 2Bx + C{x^2}}}}  = \frac{{\rm{1}}}{{\sqrt {AC - {B^{\rm{2}}}} }}\arctan \frac{{Cx + B}}{{\sqrt {AC - {B^{\rm{2}}}} }}$ for $AC > {B^{\rm{2}}}$,  and the fact that $\cos \theta  > 0$ for $\theta  \in \left( { - \frac{\pi }{2},\frac{\pi }{2}} \right)$.
Furthermore, for the extreme case when $\theta  =  \pm \frac{\pi }{2}$ and $r > \frac{{Md}}{2}$, it can be verified that the expression \eqref{approximateintegralDerivation} is still valid by taking its limit. The proof of Theorem~\ref{SNRapproximationTheorem} is thus completed.
\section{Proof of Lemma 2}
Based on the alternative expression of ${\Delta _{{\rm{span}}}}\left( M \right)$ in Lemma~\ref{angularSpanExpression}, when $r \gg \frac{{Md}}{2}$, we have
\begin{equation}\label{deltaSpanApproximation}
\small{
\begin{aligned}
&{\Delta _{{\rm{span}}}}\left( M \right) = \arctan \left( {\frac{{\frac{{Md}}{2}\cos \theta }}{{r - \frac{{Md}}{2}\sin \theta }}} \right) + \arctan \left( {\frac{{\frac{{Md}}{2}\cos \theta }}{{r + \frac{{Md}}{2}\sin \theta }}} \right)\\
&\approx 2\arctan \left( {\frac{{Md}}{{2r}}\cos \theta } \right)\mathop  \approx \limits^{\left( a \right)} \frac{{Md}}{r}\cos \theta,
\end{aligned}}
\end{equation}
where ${\left( a \right)}$ follows from the fact that $\arctan x \approx x$ for $\left| x \right| \ll 1$. By substituting \eqref{deltaSpanApproximation} into \eqref{ApproxiamtionSNR}, we have
\begin{equation}
{\gamma _{{\rm{SW}}}} \approx \bar P\frac{{{\beta _{\rm{0}}}}}{{dr\cos \theta }}\frac{{Md}}{r}\cos \theta  = \bar P\frac{{M{\beta _{\rm{0}}}}}{{{r^2}}} = {\gamma _{{\rm{UPW}}}}.
\end{equation}
This thus completes the proof of Lemma~\ref{GeneralExpressionlemma}.

 \end{appendices}

\bibliographystyle{IEEEtran}
\bibliography{ref1}

\end{document}